\documentclass[11pt]{article}

\usepackage{euscript}

\usepackage{latexsym}
\usepackage{amsmath, amssymb,graphics}
\usepackage[all]{xy}

\usepackage{fullpage}

\newtheorem{theorem}{Theorem}

\newtheorem{proposition}[theorem]{Proposition}

\renewcommand{\(}{\begin{equation*}}
\renewcommand{\)}{\end{equation*}}
\newcommand{\bea}{\begin{eqnarray*}}
\newcommand{\eea}{\end{eqnarray*}}
\newcommand{\R}{{\mathbb R}}

\newcommand{\Z}{{\mathbb Z}}
\newcommand{\Q}{{\mathbb Q}}

\newcommand{\cP}{\ensuremath{\mathcal P}}

\newcommand{\bo}{\raise-1mm\hbox{\Large$\Box$}}              
\newcommand{\beq}{\begin{equation}}
\newcommand{\eeq}{\end{equation}}

\numberwithin{equation}{section}

\renewcommand{\(}{\begin{equation}}
\renewcommand{\)}{\end{equation}}

\def\R{{\mathbb R}}
\def\Z{{\mathbb Z}}
\def\Q{{\mathbb Q}}

\def\1{{\bf 1}}

\def\<{\langle}
\def\>{\rangle}

\numberwithin{equation}{section}

\renewcommand{\(}{\begin{equation}}
\renewcommand{\)}{\end{equation}}

\begin{document}

\begin{titlepage}


\vspace{2em}
\def\thefootnote{\fnsymbol{footnote}}

\begin{center}
{\Large\bf  
Topological aspects of the 
NS5-brane}
\end{center}
\vspace{1em}

\begin{center}
Hisham Sati 
\footnote{e-mail: {\tt
hsati@pitt.edu} \newline
\newline
}
\end{center}

\begin{center}
Department of Mathematics\\
University of Pittsburgh\\
Pittsburgh, PA 15260

\end{center}


\vspace{0em}
\begin{abstract}
\noindent
In this note we investigate certain topological aspects of the 
effective action of the NS5-brane in type IIA string theory. 
 To facilitate the study  of the corresponding 
 partition function,  we define a cohomology 
class whose vanishing is a necessary condition for this function to be well-defined.
This leads to various topological conditions, including a twisted Fivebrane 
structure as well as secondary cohomology operations arising from a 
K-theoretic description. We explain how these operations
also generate the topological part of
the action as well as the phase of the partition function.
Part of the discussion also applies to  the M5-brane.

\end{abstract}

\end{titlepage}

\tableofcontents

\section{Introduction}

%
%
The NS5-brane in type IIA string theory 
is a magnetic brane \cite{CHS} which 
enjoys many interesting physical and mathematical
properties (see e.g. \cite{D}). This brane is charged under the 
dual $*H_3=H_7$ of the B-field $H_3$
and hence admits as a source 
the right hand side of the equation of
motion of the B-field. The field $H_7$, unlike $H_3$, is hence not
closed. 
This brane is obtained from the dimensional reduction of the
M5-brane from eleven to ten dimensions, and hence shares 
many of the properties of the M5-brane. 
For example, the worldvolume theory  
carries a 2-form potential as part of the chiral $(0,2)$ supermultiplet. 
Associated with the M5-brane are
several new geometric and topological structures that the author
has been investigating \cite{tcu} \cite{tcu2} \cite{tcu3} \cite{Wu}. 
This note is in some sense a continuation of that work but
for the closely related setting of the NS5-brane.

\paragraph{The action.}

An action for the NS5-brane is given in Ref. \cite{BNS}.
Another one, which is free of gravitational anomalies,
 for the coupled system of IIA $D=10$ dynamical supergravity interacting with an NS5-brane
 is constructed in \cite{CL}. We will be mostly interested in the topological terms in the
 action, as described in \cite{Wi2} \cite{W-Duality}. The main term is $\int_{M^6} H_3 \wedge C_3$,
 where $C_3$ is the Ramond-Ramond 3-form potential, reduced from the M-theory 
 C-field. We will be interested in lifting this action to seven and eight dimensions.

\paragraph{The partition function.}
The partition function of the NS5-brane is used in \cite{Wi2} \cite{W-Duality} 
essentially as a tool to study the partition function of the M5-brane. 
This is studied further in \cite{D} \cite{DVV} in the special case when the 
worldvolume is a Calabi-Yau threefold.
Instanton effects from Euclidean NS5-branes in type IIA string theory lead to
quantum corrections to the hypermultiplet moduli space in Calabi-Yau compactifications 
of type IIA string theories and are governed by the wave function of the topological B-model
\cite{APP1} \cite{APP2}. We will instead concentrate on the topological
conditions and constraints associated with the partition function on general manifolds with some structure.

\paragraph{D-branes ending on NS5-branes.}
D-branes can end on NS5-branes \cite{W-sol} \cite{Pa}.
The tension of the NS5-brane is $(2\pi)^{-5} \alpha'^{-3} g_s^{-2}$ while 
that of a D$p$-brane is $(2\pi)^{-p} \alpha'^{(-p-1)/2} g_s^{-1}$, so that 
NS5-branes are heavier than D-branes in the perturbative regime and 
the NS5-brane is a solitonic object at weak coupling (i.e. when $\alpha'$ is small). 
To study the dynamics of D-branes in the vicinity 
of NS5-branes in the weak string coupling regime, 
the NS5-brane is usually taken to be static
\cite{EGKRS} \cite{MMS2}. 
 In particular, a D2-brane ending on 
an NS5-brane in type IIA string theory 
can be obtained from dimensional reduction of a system 
consisting of an M2-brane ending on an M5-brane in M-theory. 
In fact, the NS5-brane behaves in many respects like the M5-brane, 
and so insights from one can be used to shed light on the other.
We will analyze the first system in a way that is similar to the 
analysis of the latter in \cite{DMW}. In particular, we will 
provide conditions for the NS5-brane partition function $Z_{NS5}$ to be well-defined
without going into evaluating it.
The M-branes admit a description in terms of integral cohomology
(however, more recent developments suggest generalized cohomology, see \cite{KSpin}
\cite{tcu}). For the NS5-brane, we have Ramond-Ramond fields on the 
right hand side of the equation of motion of the B-field. Thus we 
include in our analysis 
some effects arising from the K-theoretic description of these fields
\cite{FH} \cite{MW}.

\paragraph{Relating classes on the brane to classes on space-time.}
We will consider the action of the integral Steenrod square 
cohomology operations 
$Sq^i_\Z: H^m(X; \Z) \to H^{m+i}(X; \Z)$ 
(see the beginning of Section \ref{sec RR})
on cohomology classes
on the target ten-manifold $X^{10}$ of type IIA string theory. 
We will also consider the NS5-brane with 
worldvolume $M^6$, a 6-dimensional oriented manifold, as well as its extension
to a 7-manifold $Y^7$ and then to an 8-manifold $Z^8$.
All three instances of the (extended) worldvolume admit maps to 
spacetime since they represent a brane living in target space-time.  
Since the Steenrod square is natural, the following diagram is commutative
\(
\xymatrix{
H^*(W; \Z) 
\ar[rr]^-{f^*}
\ar[d]^{Sq^i}
&&
H^*(X^{10}; \Z)
\ar[d]^{Sq^i}
\\
H^*(W; \Z) 
\ar[rr]^-{f^*}
&&
H^*(X^{10}; \Z)\;,
}
\)
where $W$ stands for any of the three manifolds $M^6$, $Y^{7}$, or $Z^8$.
That is, if we have a class $x \in H^*(X^{10}; \Z)$ then puling back 
to $W$ and applying the Steenrod square gives the same result as applying the 
Steenrod square and then pulling back the result; this summarized by the 
relation $f^* Sq^i x= Sq^i f^*x$. Thus, in particular, if we take $x$ to be 
a degree four class and $i=3$, then classes on $X^{10}$ that 
are annihilated by the first differential in the 
Atiyah-Hirzebruch spectral sequence (as in \cite{DMW}) 
pull back to classes on $W$ with the same property.

\vspace{3mm}
We will apply the setting of the Chern-Simons construction 
\cite{W-flux} \cite{Wi2} 
to the NS5-brane.
That is, we
 consider the lift of the NS5-brane worldvolume $M^6$ 
via the circle to get the 7-dimensional circle bundle $Y^7$, 
which we take to be a boundary of an 8-manifold $Z^8$.
As explained in \cite{Wi2} (mostly for the similar case of the 
M5-brane), the extension to $Z^8$ can be 
used to define the theory in six dimensions. 
Most applications of the fivebrane requires the 
six-manifold to be a Calabi-Yau space 
(see e.g. \cite{D}). The spaces $Y^7$ and 
$Z^8$ should generally have compatible structures
(cf. \cite{M-framed}).

\vspace{3mm}
In section \ref{sec new} we define a new class $\Pi$, which detects
when the partition function can be well-defined. 
The class $\Pi$ is integral  for SU($n$) 
 bundles of even 
rank and even second Chern class, and its vanishing in cohomology 
is a
necessary condition for the partition function of 
the NS5-brane to be nonzero. This is equivalent to saying that space-time
admits a twisted Fivebrane structure, in the sense of \cite{SSS3}.

\vspace{3mm}
In section \ref{sec RR} we start considering the effect of 
the K-theoretic description of Ramond-Ramond fields $F_i$.
First, we show in section \ref{sec F4} that this  
 requires
the cohomology class corresponding to $F_4 \wedge F_4$ 
(see expression \eqref{eq new})
to be the mod 4 reduction of an integral class.
This uses secondary cohomology operations which are novel in this context, namely 
the Toda brackets.
Then, in section \ref{sec F26} we show that 
on an almost complex 8-dimensional manifold the cohomology classes of the 
fields $F_2$ and $F_6$ satisfy certain constraints arising from the 
Euler characteristic and the one-loop polynomial $I_8$. 

\vspace{3mm}
Some consequences of the description of the fields and the corresponding 
constraints are considered in section \ref{sec con}. The first effect, described
in section 
\ref{sec op}, is that 
the $F_4$-term in $\Pi$ when annihilated by the first differential in 
K-theory gives rise to a quadratic refinement of a bilinear form. 
The topological part of the action of the NS5-brane involving 
degree four classes 
is given by a secondary cohomology 
operation $\Omega$ on classes $z \in H^8(Z^8; \Z)$ which 
are annihilated by the first differential in the Atiyah-Hirzebruch spectral 
sequence for K-theory.
Finally, in section \ref{sec quad} we show that the
 phase of the partition function (for torsion worldvolume fields) 
 is captured by the signature $\sigma$ via $\tfrac{1}{8}\sigma(Z^8)$, 
and is the $\Z_8$-valued phase of the Pontrjagin square operation on the degree
four class. We do this in the setting of Gauss sums and cohomology
operations.

\vspace{3mm}
Due to the similarities between the type IIA NS5-brane 
and the M-theory M5-brane, a good part of our discussion 
will apply to the latter as well.

\section{The new class and its effect on the partition function}
\label{sec new}

In this section we introduce a new class and study its effect on 
the ability to define the partition function $Z_{NS5}$. 

\paragraph{Defining the new class.}
The equation of motion for the B-field is given by (see \cite{BeM} \cite{FS} \cite{MS})
\(
d(e^{-2\phi} *H_3)= F_0 \wedge F_8 
+ F_2 \wedge F_6
-\tfrac{1}{2} F_4 \wedge F_4 + I_8\;,
\label{EOM}
\)
where 
$I_8$ is the one-loop polynomial \cite{DLM} \cite{VW}
\(
I_8=\tfrac{1}{48}[p_2(X) - \lambda(X)^2]\;,
\label{I8}
\)
where $\lambda(X)=\tfrac{1}{2}p_1(X)$ is the first Spin characteristic class. 
We assume a constant dilaton $\phi$.
\footnote{This is for simplicity of the topological 
description. The physical situation requires a variable dilaton,
and restoring this is straightforward and will not change the result. }
This is a scalar field, i.e. a function $\phi: X^{10} \to \R$. 
Here $F_{2i}$ are the RR fields, viewed classically as differential forms of degree $2i$, 
and $*$ is the Hodge operator in ten dimensions. We are also interested in the 
corresponding cohomology classes, both de Rham $[-]_{\rm dR}$ and integral $[-]$,
needed for quantization.

\paragraph{Remarks.} We emphasize the following points, essential to our discussion. 
\begin{enumerate}
\item The Ramond-Ramond (RR) fields in supergravity satisfy `twisted Bianchi identities', 
that is they are annihilated by the twisted de Rham differential $d_H: d - H_3 \wedge$
rather than by the de Rham differential $d$: $dF + H_3 \wedge F=0$, where
$F=\sum_{i} F_{2i}$ is the total RR field. However, we would like to restrict to situations
where the fields are annihilated by $d$. For that, we separate the RR fields into 
those of degree $4k$ and those of degree $4k+2$ for $k=0,1$, and notice via
\(
dF_{4k}= H_3 \wedge F_{4k-2}\;, 
\qquad \qquad 
dF_{4k+2}=H_3 \wedge F_{4k}
\)
that the former type of fields appear on the RHS of the would-be Bianchi identity 
of the other and vice versa. Hence if we restrict to situations where we have either
$F_{4k}$ or $F_{4k+2}$ at a time, then each set of fields will satisfy the usual 
Bianchi identity $dF_i=0$. This can be done either via some coupling analysis as above, or 
via some finite group action, or imposed by hand. So, while the fields $F_i$ will not
be cohomology classes in general, we will consider the cases when $F_i$ can be. Furthermore, 
when they are cohomology classes, they are not always integral, but we would like to 
identify when that happens and explore some consequences.

\item The RR fields $F_i$ at the level of string theory should be 
classified (up to subtleties) by twisted K-theory, where the twist is given by the
curvature $H_3$ of the B-field (viewed as a connection on a gerbe). 
This field is a priori needed for the discussion of the 
NS5-brane. However, we would like to restrict our discussion to untwisted 
K-theory, and so we will take the cohomology class of the twist to be 
trivial $[H_3]=0$. \footnote{Again the physical situation prefers a nonzero class, but for 
simplicity we would like to work with objects within K-theory
rather than having to deal with infinite-dimensional bundles and their 
more exotic classes. From a physical point of view this amounts to neglecting 
the back-reaction of the NS5-brane on the type IIA background.}
That is, we do have a B-field, but it is flat. This allows for 
discussing the NS5-brane and yet does not require the use of twisted K-theory, 
as flatness corresponds essentially to untwisted K-theory.

\item
We will not construct the partition function of the NS5-brane, but rather provide 
some topological constructions and consistency constraints involved.
 The partition function can be defined either intrinsically
or extrinsically, as in the case of the M5-brane (see \cite{DFM}). 
The first description uses fields that are intrinsic to the brane worlvolume 
without reference (necessarily) to space-time, while the extrinsic 
description relies crucially on the embedding in spacetime, both for the
topology of the configuration as well as for the corresponding fields. 
Each of the two descriptions will involve a cohomology class, whose 
vanishing is a condition for the partition function to be well-defined. 
The two descriptions are related in the sense that the two classes
are explicitly related: The class corresponding to the intrinsic description 
can be recovered by an integration over the transverse space of the 
class corresponding to the extrinsic description. This is done explicitly for the
case of the M5-brane in \cite{DFM}, and what we do here is provide the 
analog for the NS5-brane.

\end{enumerate}

The right hand side of the equation of motion \eqref{EOM}
will lead to the class associated to the 
NS5-brane. 
\footnote{This is a strategy similar to the one employed for the M5-brane in \cite{DFM}.}
We set the cosmological constant to zero, $F_0=0$, and we define the new class corresponding to the 
right hand side by passing to de Rham cohomology
\(
\Pi := \left[
 F_2 \wedge F_6
-\tfrac{1}{2} F_4 \wedge F_4 + I_8\right]_{\rm dR}\;.
\label{eq new}
\)
This is indeed a cohomology class as the following simple calculation shows
\begin{eqnarray}
d~ \Pi &=& dF_2 \wedge F_6 + F_2 \wedge dF_6 - \tfrac{1}{2}(dF_4 \wedge F_4 + F_4 \wedge dF_4) + dI_8
\nonumber\\
&=&
0 + F_2 \wedge H_3 \wedge F_4 - \tfrac{1}{2}(H_3 \wedge F_2 \wedge F_4 + F_4 \wedge H_3 \wedge F_2) + 0
\nonumber\\
&=&
0\;,
\end{eqnarray}
where we have combined the two terms in brackets to cancel the first term.

 \vspace{3mm}
We now explore properties of this class and their effect on 
setting up the NS5-brane partition function.

\paragraph{Integrality of the class.}
We investigate whether the class $\Pi$ admits an integral lift, i.e. 
from de Rham cohomology to integral cohomology. Let us consider 
each one of the three terms in expression \eqref{eq new} separately. First, the one-loop term 
$I_8$ is always integral for Spin manifolds of even $\lambda=\tfrac{1}{2}p_1$ \cite{W-flux},
i.e. for manifolds with a so-called ``Membrane structure" \cite{tcu3}.
Then for the other two terms we use the K-theoretic 
quantization of the RR fields \cite{FH} \cite{MW}
\(
F(E)= \sqrt{\widehat{A}(X)} ~{\rm ch}(E)\;,
\)
corresponding to a
bundle
 $E$.
 \footnote{Strictly speaking, we have to use a twisted Chern character ${\rm ch}_H(E)$. However,
since we are assuming the case $[H_3]=0$, then ${\rm ch}_H(E)$ reduces to the usual Chern character
${\rm ch}(E)$.}
Considering $E$ to be an SU($n$) bundle -- for instance arising from the Chan-Paton bundle 
or from the reduction to ten dimensions of an $E_8$ bundle in eleven dimensions, as in \cite{DMW} --
we have that
$c_1(E)=0$. This then implies that the cohomology class 
corresponding to the term 
$F_2 \wedge F_6$ will be zero, since $F_2=c_1(E)$. 
Thus we still have the second term; consider the degree four RR class
\(
F_4(E)=\left[\sqrt{\widehat{A}(X)} ~{\rm ch}(E)\right]_4=-{\rm rank}(E)\cdot \tfrac{1}{48}p_1(X) + c_2(E)\;.
\)
We first assume that our ten-manifold is a product $X=Y \times \R^6$, where $Y$ is a Spin 
4-manifold. This now implies that the Pontrjagin classes of $X$ are given by those of 
$Y$ (at least modulo 2-torsion). A result of Atiyah-Hirzebrch \cite{AH} states that the 
first Pontrjagin class $p_1$ of a compact oriented differentiable manifold is a homotopy 
invariant mod 24, and in fact mod 48 if $H^2(X; \Z_2)=0$. Under these assumptions, 
we now make use of the Atiyah-Singer index theorem, which gives
 that the 
index of the Dirac operator for Spin manifolds 
is an even integer in dimension four. 
As a consequence, $\tfrac{1}{48}p_1(X)$ is an integer. This, together with the 
fact that Chern classes are integral classes, implies that 
$F_4 \in H^4(X; \Z)$. This results in the class corresponding 
to the wedge product $F_4 \wedge F_4$ being integral. 
However, this does not yet imply that the class 
of $\tfrac{1}{2} F_4 \wedge F_4$ is integral, as we are dividing by 
2. This division is of no effect if 
we require the bundles $E$ and $E'$ corresponding to 
the first $F_4$ and second $F_4$, respectively, to have 
even rank and even second Chern class.
Therefore, 
\begin{proposition}
The class $\Pi$ is integral  for SU($n$) 
bundles of even 
rank and even second Chern class.
\end{proposition}

In the general case, i.e. for ten-manifolds $X$ that are not necessarily reducible to 
4-manifolds, we assume that $\tfrac{1}{2}\lambda \in \Z$ as before so that $I_8$ is integral. 
Then we write
\(
F_4(E)= -\tfrac{1}{12}{\rm rank}(E)\cdot \tfrac{1}{2}\lambda + c_2(E)\;.
\)
Then we get the same conclusion as Proposition 1, via $F_4(E) \in 2\Z$, if we have the 
following conditions satisfied: 

(i) Membrane structure: $\tfrac{1}{2}\lambda(X) \in \Z$;

(ii) $c_2(E)$ even;

(iii) $E$ is of rank a multiple of 24.

\paragraph{Explicit expression for $\Pi$.}
The one-loop polynomial \eqref{I8}, 
together with above expressions for the quadratic RR fields,
leads to the explicit expression for $\Pi$ 
\(
\Pi=\tfrac{1}{48}\left[
p_2(X) - \left(1+ \tfrac{1}{24}r^2 \right)\lambda(X)^2 - 2r \lambda(X) c_2(E) - 24 c_2(E)^2
\right]\;,
\label{exp}
\)
where $r={\rm rank}(E)$. This can be thought 
of as the analog of the one-loop
term $I_8$, or more precisely an analog of
 the Green-Schwarz anomaly-cancellation term in heterotic 
and type I string theory \cite{GS}. The latter also contains terms
that arise from Chern classes of a gauge bundle, there  
being an $E_8 \times E_8$ bundle, as opposed to
an SU($n$) bundle. Note that in relating to eleven-dimensional 
M-theory,  we can choose to retain an $E_8$ bundle-- as in \cite{S-E8}--
 and not perform the breaking down to unitary groups as
 done in \cite{DMW}.

\paragraph{Existence of a nonzero partition function.}
In Ref. \cite{DFM} a necessary condition for the existence of a nonzero
partition function for the M5-brane in M-theory was derived. This 
decoupling from the bulk 
is essentially the condition that the class $\Theta_X(C)$ 
lifting the right hand side of the equation of motion of the C-field 
be zero. We will perform an analogous analysis for the 
NS5-brane, which is the result of the vertical dimensional 
reduction of the M5-brane down to type IIA string theory.
Instead of satisfying a quadratic refinement as the 
class $\Theta$ did, we will see that the new class enjoys index-theoretic 
properties arising from the K-theoretic quantization of the
RR fields \cite{MW} \cite{FH}. 
For the case of the NS5-brane, we require that the 
D-brane not end on it so that it is decoupled. Thus, in this sense,
the D-branes ending on the NS5-branes are the analogs of the M2-brane 
ending on the M5-brane. 
We have

\begin{proposition}
A necessary condition needed for the partition function of 
the NS5-brane to be nonzero is that the class $\Pi$ is 
zero in cohomology.
\end{proposition}

Now that we have a condition for a non-zero partition
function, we ask when such a condition is satisfied. 
We provide one general such instance. 

\paragraph{Relation to Fivebrane structures.}
When $\lambda(X)=0$, that is when we have a String structure, 
expression \eqref{exp} simplifies to 
\(
\Pi=\tfrac{1}{48}p_2(X) -\tfrac{1}{2} c_2(E)^2\;.
\)
We see that setting $\Pi=0$ is equivalent to setting 
$\tfrac{1}{48}p_2(X) -\tfrac{1}{2} c_2(E)^2$ to zero. 
The first term in this difference is essentially an obstruction to a Fivebrane structure 
\cite{SSS2} and, since the second term is
integral for even $c_2(E)$, the whole expression defines a twisted
Fivebrane structure \cite{SSS3} with the twist given 
by the integral class $\tfrac{1}{2}c_2(E)^2$. 
The twist is composite, as in \cite{tcu2}. 
Note that having a Fivebrane structure implicitly also implies that
we already have a String structure (this follows from the way 
such structures are defined via obstruction theory).
Therefore,  for even $c_2(E)$ as we have assumed so far, we have the
\begin{proposition}
(i) The vanishing of the class $\Pi$ is equivalent to 
a twisted Fivebrane structure.

\noindent (ii) A necessary condition for the non-vanishing of the NS5-brane 
partition function is that spacetime admits a  twisted Fivebrane 
structure. 
\end{proposition}
\noindent This is the analog for type IIA of the description in \cite{SSS2}
\cite{SSS3} for M-theory and heterotic string theory.

\section{Constraints on RR fields}
\label{sec RR}
We have seen that the expression \eqref{eq new} for the new class 
$\Pi$ involves a combination of total degree eight of Ramond-Ramond 
fields $F_2, F_4$, and $F_6$. 
In order to gain a better understanding of this class and 
consequences of having it, 
we will investigate constraints on such fields 
arising from K-theory and from geometric structures on the underlying space. 
These in turn will constrain $\Pi$ itself.

\medskip
Cohomology operations will appear in our discussion, namely Steenrod squares
and, later, Pontrjagin squares (see section \ref{sec op}). 
There are various forms of the former operations that we use, depending on coefficients
both on the domain and the range of these operations:
\begin{enumerate}

\item {\it The ``mod 2" Steenrod square}: $Sq^i: H^*(-; \Z_2) \to H^{*+i}(-; \Z_2)$ with 
corresponding Bockstein $Sq^1$ corresponding to $i=1$. 

\item {\it Integral-valued Steenrod square}:
$Sq^{2i+1}= \beta Sq^{2i}: H^j(-;\Z_2) \to H^{j+2i+1}(-; \Z)$ with Bockstein 
$\beta: H^j(-; \Z_2) \to H^{j+1}(-; \Z)$ corresponding to the coefficient 
sequence $0 \to \Z \buildrel{\cdot 2}\over{\longrightarrow} \Z \to \Z_2 \to 0$

\item {\it Integral Steenrod square}: $Sq^{2i+1}_\Z= \beta Sq^{2i} \rho_2: H^j(-; \Z) \to H^{j+2i+1}(-; \Z)$
corresponding to the mod 2 reduction $\rho_2: H^j(-; \Z) \to H^j(-; \Z_2)$. 

\end{enumerate}

\subsection{Conditions on $F_4$}
\label{sec F4}

We would like to consider some further K-theoretic aspects of the 
fields entering in the new class $\Pi$. 
We will first concentrate on the term in \eqref{eq new} quadratic in $F_4$. 
We will take our starting point that
the classes representing the RR fields are annihilated by the 
first differential in the Atiyah-Hirzebruch spectral sequence for K-theory;
that is we take the class $x_4$ representing $F_4$ to satisfy 
$Sq^3_\Z x_4=0$. 

\vspace{3mm}
One might ask whether the above condition is all what one should impose.
Indeed, this is not the end of the story. In addition to the above primary
operation, we will have a secondary operation, valid once the 
primary one is satisfied. This secondary operation is given by the
Toda bracket 
$
\langle Sq^3_\Z, Sq^3_\Z, x_4  \rangle$.
This will in fact take us beyond the $\Z_2$ and  $\Z_3$ coefficients 
encountered in previous literature (see \cite{DMW} and \cite{ES}, respectively);
we will be dealing with $\Z_4$ coefficients.

\paragraph{Secondary cohomology operations.}
If the cohomology class $x$ is such $Sq^3_\Z x=0$ then there is a secondary 
cohomology operation that can arise and that takes the form 
\(
\langle Sq^3_\Z, Sq^3_\Z, x\rangle = \beta_2^\Z Sq^4 x\;.
\)
This can be seen as follows. 
Consider the diagram (part of a Postnikov tower)  
which depicts the various cohomology operations 
and maps that we have
\(
\xymatrix{
&
 H^{i+2}(-; \Z) 
\ar[r]
~\ar@/^1pc/[rrr]^{Sq^3_\Z}
&
X_1
\ar[d]
\ar[rr]^{\hspace{-5mm}b}
&&
H^{i+5}(-; \Z)
\\
 M 
\ar[rr]^x
\ar@{-->}[urr]^a
 \ar@{-->}[ur]
&&
H^i(-; \Z)
\ar[rr]^{Sq^3_\Z}
&&
H^{i+3}(-; \Z)\;.
}
\)
Starting with a cohomology class 
$x$ on $M$ and applying $Sq^3_\Z$ to it gives a class $Sq^3_\Z x \in  H^{i+3}(-; \Z)$.
The composition $a \circ b$ is the secondary cohomology 
operation represented by the Toda bracket 
$\langle Sq^3_\Z, Sq^3_\Z, x\rangle$.

\vspace{3mm}
We have the Adem relation $Sq^2 Sq^3=Sq^5 + Sq^4 Sq^1$, which is equal to 
$Sq^5$, applied to an integral class. 
Corresponding to the reduction mod 4
 sequence
of coefficients
 $\Z  \buildrel{\cdot 4}\over{\longrightarrow} \Z 
 \buildrel{\rho_4}\over{\longrightarrow} \Z_4$, we have the 
 following relation between the corresponding cohomology 
 groups
\(
\xymatrix{
H^i(-; \Z_4)
\ar[rr]^{\beta_2^\Z}
\ar[drr]_{\beta_2}
&&
 H^{i+1}(-;  \Z)
\ar[d]^{\rho_2}
\\
&&
H^{i+1}(-; \Z_2)\;,
}
\)
so that $\beta_2^\Z$ is the Bockstein taking cohomology in 
$\Z_4$ coefficients to integral cohomology in one 
higher degree. 
The Toda bracket then takes the form
\footnote{
The spectral sequence is 
summarized as follows

\(
\begin{tabular}{|c|}
\hline
{
\def\objectstyle{\scriptstyle}
\def\labelstyle{\scriptstyle}
\xymatrix{
&
&
\iota_2
\ar[dr]
&
&
Sq^2\iota_2 \ar[dr]
&
Sq^3\iota_2=\beta_2 Sq^4\iota
\\
\iota
&
&
Sq^2\iota
&
Sq^3\iota
&
Sq^4\iota
&
Sq^5\iota
\\
0&1&2&3&4&5
}
}
\\
\hline
\end{tabular}
\nonumber
\)
Here $\iota$ and $\iota_2$ are the generators in the indicated degrees.
}
\(
\langle Sq^3_\Z, Sq^3_\Z, x_4 \rangle = \beta_2^\Z Sq^4 x_4= \beta_2^\Z(x_4 \cup x_4)\;.
\)
Therefore, we have 
\begin{proposition}
The K-theoretic description of Ramond-Ramond fields requires
the cohomology class corresponding to $F_4 \wedge F_4$ 
to be the mod 4 reduction of an integral class.
\label{Prop mod 4}
\end{proposition}

\subsection{Conditions on $F_2$ and $F_6$}
\label{sec F26}

We will apply the setting of the 
Chern-Simons construction to the NS5-brane.
That is, we
 consider the lift of the NS5-brane worldvolume $M^6$ 
via the circle to get the 7-dimensional circle bundle $Y^7$, 
which we take to be a boundary of an 8-manifold $Z^8$.
As explained in \cite{Wi2} (mostly for the similar case of the 
M5-brane), the extension to $Z^8$ can be 
used to define the theory in six dimensions. 
The spaces $Y^7$ and 
$Z^8$ should generally have compatible structures (see \cite{M-framed}). 
For example, many applications of the fivebrane requires the 
six-manifold to be a Calabi-Yau manifold
(see e.g. \cite{D}). 
We will work a bit more generally with 8-manifolds which admit
an almost complex structure (and not necessarily Spin)
and ask for constraints on 
$F_2$ and $F_6$ due to this. This discussion is in 
spirit (although not in techniques) analogous to the
discussion of the one-loop term in \cite{IPW} \cite{BB}. 
We will use the cohomology class of $F_2$ to define a complex line bundle over 
the NS5-brane worldvolume.

\paragraph{Almost complex structures.}
An almost complex structure (acs) on a manifold $M$ of dimension $2n$
is an endomorphism $J$
of the tangent bundle $TM$ satisfying $J^2=-1$. This gives $TM$ the 
structure of a complex vector bundle. From obstruction theory, an almost
complex structure can be viewed as a section of the ${\rm SO}(2n)/{\rm U}(n)$ bundle 
associated to $TM$ (see \cite{St}). 
On the other hand, a {\it stable} almost complex structure (stable acs)
on $M$ is a reduction of the structure group of the stable tangent bundle
of $M$ from SO to U.
The manifold $Z^8$ has a stable acs $\omega$
if the tangent  bundle $TZ^8$ is stably isomorphic to the underlying 
real bundle of some complex vector bundle over $Z^8$. 
This will be an actual acs if $c_4(\omega)=\chi (Z^8)$.
The necessary conditions for the existence of a stable acs are the existence of
integral lifts (for suitable $i$) 
$c_i \in H^{2i}(M; \Z)$ of the even Stiefel-Whitney classes
$w_{2i}\in H^{2i}(M; \Z_2)$, that is $w_{2i}(M)=\rho_2c_i$, where 
$\rho_2$ is mod 2 reduction. This is equivalent to saying that the 
integral Stiefel-Whitney class in odd degree vanishes $W_{2i+1}(M)=0 \in H^{2i+1}(M; \Z)$.
Such conditions are natural to impose on branes. 
In the 8-dimensional case, the two conditions are $W_3=0$ and $W_7=0$.
The latter 
is automatically satisfied on our manifolds of dimension less than or equal to 
eight \cite{Mas}.

\vspace{3mm}
We will take $u \in H^2(Z^8; \Z)$ and $v \in H^6(Z^8; \Z)$  to
be the cohomology classes $[F_2]$ and $[F_4]$
 corresponding to the fields $F_2$ and 
$F_6$, respectively.  
Consider $Z^8$ to be an 8-manifold with a stable almost complex structure 
$\omega$. Let
$(u, v)$ to be a pair of class in $H^2(Z^8; \Z) \times H^6(Z^8; \Z)$
such that $\rho_2 u= w_2(Z^8)$ and $\rho_2 v= w_6(Z^8)$.
From the expression of the Chern character 
${\rm ch}_3=\tfrac{1}{6}(3c_3 - 3c_1c_2 + c_1^3)$ we see that 
if $c_2(\omega)=0=p_1(Z^8)$ and 
$3c_3(\omega)=c_1(\omega)^3$, then simply $u=F_2=c_1(\omega)$ 
and $v=F_3=c_3(\omega)$. In this case, the Chern classes satisfy 
\bea
2c_4(\omega) + c_1(\omega) \cdot c_3(\omega) &\equiv & 0 ~({\rm mod}~ 4)\;,
\\
2\chi(Z^8) + c_1(\omega) \cdot c_3(\omega) &\equiv & 0 ~({\rm mod}~ 4)\;.
\eea
This follows from a more general formula (see \cite{Hir}), using
$\rho_2 \chi(Z^8)= \rho_2 c_4(Z^8)= w_8(Z^8)$.
Indeed, we will apply the more general main result of \cite{He}; 
the manifold $Z^8$ has an almost complex structure if and only
if 
\begin{enumerate}
\item $w_8(Z^8) \in Sq^2 H^6(Z^8; \Z)$;
\item $\rho_2 u= w_2 (Z^8)$ and $\rho_2 v= w_6 (Z^8)$;
\item $2 \chi (Z^8) + u \cdot v\equiv 0~(\mod 4)$;
\item $8 \chi (Z^8) =
4 p_2(Z^8) - p_1(Z^8)^2 + 8 u \cdot v - u^4 + 2u^2p_1(Z^8)$ 

\end{enumerate}

\vspace{3mm}
The above discussion immediately gives the following constraint on 
the cohomology classes $u$ and $v$ corresponding to the fields $F_2$ and $F_6$,
respectively. 

\begin{proposition}
For $Z^8$  an almost complex 8-dimensional manifold, the cohomology classes 
$u$ and $v$
of the 
fields $F_2$ and $F_6$, respectively,  satisfy
\(
u\cdot v -  \tfrac{1}{8}u^4 + \tfrac{1}{2}u^2 \cdot \lambda(Z^8) = \chi(Z^8) - 24I_8(Z^8)\;. 
\)
\end{proposition}
In the particular case where the Pontrjagin classes vanish, and hence the one-loop
polynomial is zero in cohomology, the constraint simplifies to 
\(
u\cdot v = \tfrac{1}{8}u^4\;,
\)
i.e. the first term in the expression \eqref{eq new} can be traded 
for the class arising from $F_2$ only. This, together with Prop. 
\ref{Prop mod 4}, provides considerable constraints on the class $\Pi$.

\section{Consequences for the action and partition function}
\label{sec con}
In this section we consider some topological consequences of the 
classes considered in previous sections,
concentrating mostly on effects
related to the degree four field $F_4$ and its corresponding cohomology 
class.

\subsection{The action via cohomology operations}
\label{sec op}

We  will now show that the topological action involving the cohomology class of the 
degree four field $F_4$ 
is generated by cohomology operations and leads to a quadratic refinement of 
a bilinear form. We will make use of some of the constructions in \cite{CV} and 
\cite{Th}.

\vspace{3mm}
On the integral middle cohomology $H^4(Z^8; \Z)$ we have the relation among Steenrod squares
\(
(Sq^2 \rho_2) \circ (\beta Sq^2 \rho_2)=0\;,
\label{eq rel}
\)
 where $\beta$ is the 
Bockstein operation.
This can be seen as follows. For a degree four class 
$z$, we have $Sq^2 \rho_2 (\beta Sq^2 \rho_2 z)=Sq^2 Sq^1 Sq^2 \rho_2 z$, 
which gives $Sq^2 Sq^3 \rho_2 z$ via  the Adem relation 
$Sq^3=Sq^1 Sq^2$. Another version of the Adem relation leads to
$Sq^1 Sq^4 \rho_2 z + Sq^4 Sq^1 \rho_2 z$, with the second summand
being zero since $Sq^1 \rho_2=0$. Because $Sq^i$ acts as squaring 
on classes of degree $i$, 
this then gives $Sq^1 \rho_2 (z^2)=0$, again due to $Sq^1 \rho_2$ being
identically zero.

\vspace{3mm}
Let $\Omega$ denote the secondary cohomology operation associated with the
relation \eqref{eq rel}. The operation is defined on the subgroup
\(
G(\Omega, Z^8)=\left\{ z \in H^4(Z^8; \Z)~|~ \beta Sq^2 \rho_2 z=0\right\}
\)
and takes values in the quotient group $H^8(Z^8; \Z_2)/Sq^2 \rho_2 H^6(Z^8; \Z)$,
where the quotient is by the indeterminacy. 
The operation $\Omega$ is not uniquely specified by the above relation:
$\Omega'= \Omega + Sq^4$ is another secondary cohomology operation 
associated with the same relation.
One can remove this ambiguity by normalizing, i.e. by taking
$\rho_2 z^2 \in \Omega (z)$ for $z \in H^4(\mathbb{H}P^2; \Z)$,
where $\mathbb{H}P^2$ is the quaternionic projective
plane.   
We now take $z, z' \in H^4(Z^8; \Z)$ be degree four classes in the domain of 
$\Omega$ that are annihilated by $Sq^3$. Then, from \cite{Th}, the
operation $\Omega$ leads to a quadratic refinement 
  \(
  \Omega (z + z') = \Omega (z) + \Omega (z') + \{ z \cup z'\}\;,
  \)
where $\{z \cup z'\}$ denotes the image of $\rho_2(z \cup z')$ 
in $H^8(Z^8; \Z_2)/Sq^2 \rho_2 H^6(Z^8; \Z)$.

\vspace{3mm}
The following is analogous to the corresponding statement in M-theory \cite{KSpin}.
\begin{proposition}
The $F_4$-term cohomology class in $\Pi$ if annihilated by the first differential in 
K-theory gives rise to a quadratic refinement of a bilinear form. 
\end{proposition}

\paragraph{The case when $Z^8$ is Spin.} 
When $Z^8$ is taken to be Spin, then
for all $z \in H^4(Z^8; \Z)$ the action of 
the operation of mod 2 reduction on the expression appearing in the 
topological part of the action leads to
\bea
\rho_2( zq_1(Z^8) - z^2) &=& w_4(Z^8) \rho_2 z - \rho_2 z^2
\nonumber \\
&=& Sq^4 \rho_2 z - \rho_2 z^2\;,
\eea
which gives zero. Since $H^8(Z^8; \Z)\cong \Z$, there is 
a unique $y \in H^8(Z^8; \Z)$ such that 
$2y= z\lambda(Z^8) - z^2$. Then it makes sense to 
denote $y=\tfrac{1}{2}(z\lambda(Z^8) - z^2)$.
 In the Spin case,
for every $z \in G(\Omega, Z^8)$ we have 
\(
\Omega (z) = \rho_2 \tfrac{1}{2}\left( z \lambda (Z^8) - z^2\right)\;,
\)
which is the topological part of the action, in the M5-brane case
\cite{Wi2} \cite{HS}.
Therefore, 
\footnote{While this is done for the case when $Z^8$ has no boundary, we 
will consider the more general case in the next section.}
\begin{proposition}
The topological part of the action of the NS5-brane involving 
degree four classes 
is given by the secondary cohomology 
operation $\Omega$ on classes $z \in H^8(Z^8; \Z)$ which 
are annihilated by the first differential in the Atiyah-Hirzebruch spectral 
sequence for K-theory.
\end{proposition}

\subsection{Quadratic refinements and the phase of the partition function}
\label{sec quad}

Having identified various structures 
needed in order to properly describe 
 the topological 
action, we now turn to effects on the partition function on which we consider 
the corresponding topological effects. 
Using this, we will show that the phase of the partition function,
given by one eighth of the signature of the 8-manifold, is
captured by a secondary cohomology operation.  
We will describe the phase of the partition function 
via the Pontrjagin square operation as well as (equivalently) 
via Gauss sums. 
The discussion in this section also applies to the M5-brane.
In both cases, we have the topological action 
taking the form $\tfrac{1}{2}\int_{M^6} H_3 \wedge C_3$,
$\tfrac{1}{2}\int_{Y^7} G_4 \wedge C_3$,
and $\tfrac{1}{2}\int_{Z^8}G_4 \wedge G_4$
in six, seven, and eight dimensions. Here $G_4$ refers to the RR field
$F_4$ for the case of the NS5-brane, while it is the C-field field strength in 
M-theory for the case of the M5-brane.

\paragraph{Reduction of the Hirzebruch L-polynomial.}
The topological action will involve the Hirzebruch L-polynomial 
of $Z^8$, via $\tfrac{1}{8}L_2(Z^8)$ as in \cite{HS}. We would like to characterize 
the division by 8. The division by 2 and 4 are easy to describe 
via reduction mod 2 and mod 4, respectively. 
Indeed, for instance from the  formulas in \cite{Th2} we have 
 \bea
\rho_2L_2&=& \rho_2 (\cP(v_4))= v_4^2 \quad  \in H^8(BSO; \Z_2)\;.
\\
\rho_4 L_2 &=& \cP(v_4) + i_*(w_8 + w_6w_2 + w_2^4) \quad \in H^8(BSO; \Z_4)\;,
\eea
Here $w_i$ and $v_i$ are the Steifel-Whitney class of degree $i$ and 
the Wu class of degree $i$, respectively, 
$\mathcal{P}$ is the Pontrjagin square cohomology operation
\(
\mathcal{P}: H^4(Z^8; \Z_2) \longrightarrow \Z_4\;,
\label{eq beta}
\)
and $i: \Z_2 \hookrightarrow \Z_4$ is the inclusion.
What we need is reduction modulo 8. We study this using the 
classic constructions in \cite{BM}.

\vspace{3mm}
One main point here is that the topological part of the 
 partition function in our context can be viewed 
 essentially as an instance of a Gauss sum.
 A special role will be played by the torsion part of the fields.  
 We consider a finite group $T$, which will be a relative cohomology
 group with $\Z_2$-coefficients when the worldvolume field $F_4$ 
  is torsion-free or a torsion subgroup in the torsion case. 
 We will make use of the results of Ref. \cite{Ta} in what follows.

\paragraph{Gauss sums.}
The action will be a function $S: T \to \R/\Z$. The associated
Gauss sum corresponding to $S$ is given by
\(
\mathcal{G}(S)= \sum_{x \in T} e^{2\pi i S (x)}\;.
\)
This expression splits into a magnitude or norm, $N(S)=|\mathcal{G}(S)|$, and a 
phase $\beta (S)\in \R/\Z$ (which can be defined only if the norm is 
nonzero) via
\(
\mathcal{G}(S)= N(S) \cdot e^{2\pi i \beta (S)}\;.
\)
Let $K$ be a subgroup of $T$ the restriction of the action $S$ to which is 
zero, i.e. $S|_K=0$. Then it can be shown (see
\cite{Ta}) that the Gauss sum is determined by the restriction to the complement
\(
\mathcal{G}(S)= |K| \cdot \mathcal{G}(S|_{K^\perp/K})\;,
\label{eq perp}
\)
which is clear on physical grounds.
Note that $S_{K^\perp/K}: K^{\perp}/K \to \Q/\Z$ is well-defined 
since $K \subset K^\perp$.

\paragraph{The phase when $Z^8$ closed.}
Consider the case when $Z^8$ is closed, in addition to being 
connected and oriented. 
The cup product pairing 
$
H^4(Z^8; \Z) \times H^4(Z^8; \Z) \to \Z_2
$
admits a quadratic refinement given by the Pontrjagin square
$\mathcal{P}$.
By Morita's proof \cite{Mor} of Brown's conjecture, the  phase of the 
Pontrjagin square is given by 
\(
\beta (\mathcal{P})=\tfrac{1}{8}\sigma(Z^8) \in \Q/\Z\;,
\)
where $\sigma(Z^8)$ is the signature of the 8-manifold $Z^8$.

\paragraph{The phase when $Z^8$ has boundary.}
We work with relative cohomology of the pair $(Z^8, Y^7)$, where 
$Y^7$ is the boundary of $Z^8$.
In this case, the Pontrjagin square $\mathcal{P}: H^4(Z^8, Y^7; \Z_2) \to \Z_4$
is still the quadratic refinement of the cup product pairing. 
However, in this case we distinguish two cases, depending on whether or not
the cohomology of $Y^7$ has torsion in degree four. 
That is, whether or not the worldvolume fieldstrength 
$G_4$ is torsion. 

\paragraph{1.}{\it Torsion-free $G_4$.}
If $H^4(Y^7; \Z)$ 
is torsion-free, then \eqref{eq beta} holds.
This can be seen as follows (see \cite{Ta} for a general discussion).  
When there is a boundary, the cup product pairing has an annihilator:
if $T=H^4(Z^8, Y^7; \Z_2)$, then $T^\perp$ is the image of $H^3(Y^7; \Z_2)$
in $H^4(Z^8, Y^7; \Z_2)$.
Let us start with a worldvolume degree three cohomology class $x\in H^3(Y^7; \Z_2)$
and get a $\Z_4$ quantity in two different ways. 
On the one hand, from the cohomology exact sequence corresponding to the relative
pair ($Z^8$, $Y^7$), we can map $x$ to a class in $H^4(Z^8, Y^7; \Z_2)$,
on which we can apply the Pontrjagin square to get 
\(
H^3(Y^7; \Z_2) \to H^4(Z^8, Y^7; \Z_2) 
\buildrel{\mathcal{P}}\over\longrightarrow 
H^8(Z^8, Y^7; \Z_4)\cong \Z_4\;.
\label{comp 1}
\)
On the other hand, starting with $x$, we take its cup product with 
the degree four class $Sq^1 x$ to get a class in $H^7(Y^7; \Z_2)$
\(
\xymatrix{
H^3(Y^7; \Z_2) 
\ar[rr]^{\hspace{-1cm}x \cup Sq^1 x}
&&
H^7(Y^7; \Z_2) \cong \Z_2 \subset \Z_4
}\;.
\label{comp 2}
\)
It follows from the general results of \cite{TH} that the two compositions,
\eqref{comp 1} and \eqref{comp 2},
are equal. 
If $H^4(Y^7; \Z)$ is torsion-free then $Sq^1 x=0$, so that $\mathcal{P}$ 
vanishes on $T^\perp$. We can identify $T/T^\perp$ with the image of the relative
cohomology
$H^4(Z^8, Y^7; \Z_2)$ in $H^{4}(Z^8; \Z_2)$. 


\vspace{3mm}
The enhancement $\mathcal{P}$ is related to the mod 2 reduction of the integral form.
If there is no 2-torsion in $H^4(Y^7; \Z)$ then the image of $H^4(Z^8, Y^7; \Z)$ in 
$H^4(Z^8; \Z)/{\rm torsion}$ has a nonsingular bilinear pairing with a matrix
$B_Z$ of determinant ${\rm det}B_Z=\pm 1$. 
This gives that $\mathcal{P}$ and $S_{B_Z}$ are equivalent.

\paragraph{2.} {\it Torsion $G_4$.} We now consider the case when $H^4(Y^7; \Z)$ has torsion,
with torsion subgroup $T^4(Y^7)$. For $i: Y^7 \hookrightarrow Z^8$ the inclusion
and $j: (Z^8, Y^7) \to Z^8$ the map that forgets the boundary, 
the various cohomology groups are related via the induced
maps $i^*: T^4(Z^8) \to T^4(Y^7)$
and $j^*: H^4(Z^8, Y^7; \Z) \to H^4(Z^8; \Z)$.
We take the torsion part of the action in seven dimensions 
 $S: T^4(Y^7) \to \Q/\Z$ to be a quadratic function over the linking pairing
$L: T^4(Y^7) \times T^4(Y^7) \to \Q/\Z$. 
Let $\hat{v}_4 \in H^4(Z^8, Y^7; \Z_2)$ and $v_4\in H^4(Z^8; \Z)$
be liftings of the Wu class $v_4\in H^4(Z^8; \Z_2)$ 
to relative mod 2 cohomology and integral absolute cohomology, respectively, 
 compatible with $S$. 
Compatibility with $S$ 
means that for all $x \in T^4(Y^7)$ such that $x=i^*(y)$ where
$y\in H^4(Z^8; \Z)$, we have 
\(
S(x)=\tfrac{1}{2}\langle y \cdot \hat{v}_4, [Z^8, Y^7]\rangle 
-\tfrac{1}{2}\langle y \cdot (j^*)^{-1}y, [Z^8, Y^7] \rangle \in \Q/\Z\;.
\)
This is a mathematical statement corresponding essentially to the physical requirement of 
consistent dimensional reduction and lifting, and it follows from \cite{BM}
that this is always the case. 
As a consequence, we have
$j^*(\hat{v}_4)=\rho_2(v_4')$;
 there is a class $b \in T^4(Y^7)$ such that the corresponding induced action on $Y^7$,
$S i(x)= L(b, i(x))$, for 
$x\in T^4(Z^8)$ and this is related to a class $t \in T^4(Z^8)$ on $Z^8$ such that $i^*(t)=2b$.
This allows us to divide the action by 2, as in \cite{W-flux} \cite{Wi2}.

\vspace{3mm}
For $b \in H^4(Y^7; \Z)$ a class on the seven-dimensional worldvolume and $z \in H^4(Z^8, Y^7; \Z_2)$ 
a class that is trivial on that worldvolume, we have the pairing 
$
\langle \cP(z), [Z^8, Y^7] \rangle =L(b,b) \in \Z_4$.
Now the index is given by
$
I(Z^8)=\langle v_4^2, [Z^8, Y^7]\rangle + L(b,b) - |\mathcal{G}(T^4(Y^7), S)| \in \Z_4
$,
where $|\mathcal{G}(T^4(Y^7), S)|$ is the norm of the Gauss sum of the action $S$ on the 
torsion subgroup $T^4(Y^7)$ of the seven-dimensional worldvolume. 
Therefore,  the signature mod 8 is given by the formula
\(
\sigma (Z^8)= \left( \langle v_4^2, [Z^8, Y^7]\rangle 
+ S(b)\right) - |\mathcal{G}(T^4(Y^7), S)| \in \Z_8\;,
\label{eq main}
\)
We summarize our discussion above with the following

\begin{proposition}
The phase of the partition function, given by $\tfrac{1}{8}\sigma(Z^8)$, 
can be described as the phase of the Pontrjagin square operation on the degree
four class. In the case when $H^4(Y^7; \Z)$ has torsion, the expression for the phase 
is given by \eqref{eq main}.
\end{proposition}

The first part of the statement can be viewed as another way of viewing the 
phase than the ones described in \cite{Wi2} \cite{HS}, for the case when the Wu class can be set to 
zero (see \cite{Wu} for a discussion of when this can occur). 
Other effects of the Pontjagin square in the case of the M5-brane is 
considered in \cite{tcu3}.
The general index problem in the presence of a boundary can be described 
in a way analogous to the discussion in \cite{S-sig} for the case of M-theory.

\vspace{0.5cm}
\noindent {\large \bf Acknowledgements}

\vspace{2mm}
The author  would like to thank Mike Hill for useful discussions on
secondary cohomology operations. 
He also acknowledges the kind hospitality of 
the American Institute of Mathematics, Palo Alto, and
IHES, Bures-sur-Yvette, during the work on this project. 
 This research is supported by NSF Grant PHY-1102218.


\end{document}